\documentstyle[prl,preprint,eqsecnum,aps,epsf]{revtex}
\begin{document}

\title{Newton Law on the Generalized Singular Brane with and without $4d$
Induced Gravity} 
\author{Eylee Jung\footnote{Email:eylee@mail.kyungnam.ac.kr}, SungHoon Kim\footnote{Email:shoon@mail.kyungnam.ac.kr}, and
D. K. Park\footnote{Email:dkpark@hep.kyungnam.ac.kr 
}, }
\address{Department of Physics, Kyungnam University, Masan, 631-701, Korea}

\maketitle

\maketitle
\begin{abstract}
Newton law arising due to the gravity localized on the general singular
brane embedded in $AdS_5$ bulk is examined in the absence or presence of 
the $4d$ induced Einstein term. For the RS brane, apart from the subleading
correction, Newton potential obeys $4d$-type and $5d$-type gravitational
law at long- and short-ranges if it were not for the induced Einstein 
term. The $4d$ induced Einstein term generates an intermediate range at
short distance, in which the $5d$ Newton potential $1/r^2$ emerges. For
Neumann brane the long-range behavior of Newton potential is exponentially
suppressed regardless of the existence of the induced Einstein term. For 
Dirichlet brane the expression of Newton potential is dependent on the 
renormalized coupling constant $v^{ren}$. At particular value of $v^{ren}$
Newton potential on Dirichlet brane exhibits a similar behavior to that on
RS brane. For other values the long-range behavior of Newton potential is 
exponentially suppressed as that in Neumann brane.

\end{abstract}

\newpage
\section{Introduction}
A braneworld scenario is a physical picture which assumes that our $4d$ 
spacetime universe is embedded in higher-dimensional world. Although 
higher-dimensional theory 
has its own long history\cite{lee84,ruba83,viss85} in the context of 
Kaluza-Klein theory or not, modern braneworld scenarios such as 
large extra dimensions\cite{ark98-1,anto98} or warped extra 
dimensions\cite{rs99-1,rs99-2} seem to be mainly motivated from the 
string theories\cite{hora96}. Making use of the recent scenarios a flurry 
of activities has tried to examine the various physical problems such as 
big bang universe\cite{bine99,csa99,cline99}, cosmological constant 
hierarchy\cite{kim01,alex01,park01-1}, brane inflation\cite{ark03-1}, 
and black hole physics\cite{cham00,emp00,gidd00,gidd01-1,gidd02-1,gidd02-2}. 
In this paper we will examine Newton law arising due to the gravity localized
on the brane when general singular brane is embedded in $AdS_5$ bulk. 

The Newton law on the brane was firstly computed by RS in Ref.\cite{rs99-2}
when the bulk spacetime is $AdS_5$. In this case the gravity localized on
the brane yields an $4d$-type $1/r$ and $1/r^3$ subleading correction arising 
due to Kaluza-Klein exictation:
\begin{equation}
\label{rs-newton}
V \sim G \frac{m_1 m_2}{r} 
\left(1 + \frac{R^2}{r^2} \right)
\end{equation}
where $R$ is a radius of $AdS_5$. Under the same setup an improvement of the
potential was tried by involving the brane-bending effect\cite{gidd00,garr00}
and by computing the one-loop correction to the gravitational 
propagator\cite{duff00}, which slightly changes the sub-leading correction
by a overall multiplication factor
\begin{equation}
\label{imp-newton}
V \sim G \frac{m_1 m_2}{r}
\left( 1 + \frac{2}{3} \frac{R^2}{r^2} \right).
\end{equation}

Subsequently, the gravitational propagator for the linearized fluctuation 
equation in RS picture is generally computed by adopting the singular
quantum mechanics(SQM) as a calculational tool\cite{park02-1,park02-2}. 
From the viewpoint of SQM the linearized gravitational propagator is 
crucially dependent on the boundary condition(BC) that the propagator
should satisfy at the location of the brane. The physically relevant BC
is not uniquely determined in the framework of SQM due to the singular 
nature of the potential and the physically consistent BCs are parametrized 
by a real
parameter, say $\xi$. The physically consistent BCs are in general introduced
by a self-adjoint extension technique\cite{cap85,reed75}, which is effectively
identical to the coupling constant renormalization\cite{jack91,park95}.  
Recently, this generalized gravitational propagators are used to compute the
Newton potential on the brane\cite{park03-1,park03-2} with a particular
choice of $\xi$ as $\xi = 1/2$, which means the Dirichlet and Neumann BCs
are included with an equal weight.

Recently, the Newton law with different setup is also considered. Especially,
much attention is paid to the case of Minkowski bulk with an involving the 
$4d$ induced Einstein term arising due to the quantum effect of one 
loop\cite{dvali00,dvali01}. In the case of the flat bulk the Newton potential
becomes $4d$-type $1/r$ at the short range, {\it i.e.} $r << \lambda$, and 
$5d$-type $1/r^2$ at the long range, {\it i.e.} $r >> \lambda$, where
$\lambda$ is a ratio of $4d$ Planck scale with that of $5d$;
$\lambda \equiv M_4^2 / M_5^3$. This fact is used to explain the observed
acceleration of the Universe\cite{ruba03}.

The effect of the induced Einstein term is also examined when the $5d$ bulk 
is $AdS_5$ spacetime\cite{kiri02,ito02}. In this setup the $5d$-type Newton
potential arises in the region of $\lambda << r << R$. At other ranges the 
$4d$ gravitational potential is recovered. Recently, this fact is again 
confirmed in the context of SQM\cite{park03-2}.

In this paper we will examine the Newton potential using the generalized 
gravitational propagators derived using SQM. Thus, the gravittational 
propagator contains in general the parameter $\xi$, which parametrizes 
the physically relevant BCs. The $\xi$-dependence of the Newton potential is 
examined throughout the paper. We will confine ourselves to the case of single
copy of the $AdS_5$ bulk. Thus our computation is limited to the original
RS plus $AdS$/CFT\cite{mal98,aharo00} setup with and without the induced
Einstein term. In particular, we will examine in detail the cases of 
$\xi = 0$, $\xi = 1$, and $\xi = 1/2$. When $\xi = 1$, we have a Neumann
brane in which the gravity acquires mass\cite{park02-1,park02-2}. It is 
well-known that the Newton potential generated due to the exchange of a
massive graviton is exponentially suppressed at long range. This fact will 
be explicitly shown in the following. At $\xi =0$ we have a Dirichlet brane. 
In this case the non-trivial gravitational propagator can be derived 
{\it via} the coupling constant renormalization\cite{park02-1,park02-2}. Thus
the final expression of the Newton potential is dependent on the 
renormalized coupling constant. We will show in the following that when the 
renormalized coupling constant has a particular valus, the Newton potential
on the Dirichlet brane is proportional to that on RS brane which corresponds
to the brane at $\xi = 1/2$. 

The paper is origanized as following. In Sec.II we will derive a formula
which shows how to compute Newton potential arising due to a localized gravity
from the fixed-energy amplitude. Using the formula Newton law in general
singular brane is examined in Sec.III without consideration of the 
$4d$ induced Einstein term when bulk is a single copy of $AdS_5$ spacetime. 
A particular 
attention is paid to the RS-brane($\xi = 1/2$), Neumann brane($\xi = 1$) and
Dirichlet brane($\xi = 0$). The effect of the $4d$ induced Einstein term to 
Newton potential is studied in Sec.IV. In Sec.V brief conclusion and further
remark are given.
In appendix A and B Newton laws on Neumann and Dirichlet branes are explicitly
derived in the presence of the $4d$ induced Einstein term.

\section{Newton Potential from the Fixed-Energy Amplitude}
In this section we will consider briefly how to compute Newton potential
arising due to the localized gravity on the brane from the fixed-energy 
amplitude\footnote{The definition of the fixed-energy amplitude is a Laplace
transform of the Euclidean propagator.} computed in the context of the 
SQM. As an example we will consider the usual RS scenario without the induced
Einstein term. However, the final result is model-independent.

Let us consider the $5d$ Einstein equation\footnote{Our notation is that 
$M$ and $N$ are bulk spacetime indices and, $\mu$ and $\nu$ are brane spacetime
indices.}
\begin{equation}
\label{eineq-1}
R_{MN} - \frac{1}{2} G_{MN} R = 
-\frac{1}{4M^3}
\left[ \Lambda G_{MN} + v_b G_{\mu \nu} \delta_M^{\mu} \delta_N^{\nu}
       \delta(y) \right]
\end{equation}
where $M$, $\Lambda$ and $v_b$ are $5d$ Planck scale, $5d$ cosmological 
constant and brane tension
respectively. As is well-known it has a solution
\begin{equation}
\label{rs-sol1}
ds^2 = e^{-2 k |y|} \eta_{\mu \nu} dx^{\mu} dx^{\nu} + dy^2
\end{equation}
if the fine-tuning conditions $\Lambda = -24M^3 k^2$ and $v_b = 24M^3 k$ are 
satisfied.

Introducing the linearized gravitational fluctuation $h_{\mu \nu}(x, y)$ as 
following
\begin{equation}
\label{rs-fluc1}
ds^2 = \left[ e^{-2k|y|} \eta_{\mu \nu} + h_{\mu \nu}(x, y) \right]
dx^{\mu} dx^{\nu} + dy^2
\end{equation}
and ignoring the tensor structure for simplicity by choosing a gauge
$h_{\mu \nu, \mu} = h_{\mu}^{\mu} = 0$, one can derive the following 
linearized fluctuation equation
\begin{equation}
\label{dasi1}
\left[ e^{2 k |y|} \Box^{(4)} + \partial_y^2 - 4 k^2 + 4 k \delta(y) \right]
h_{\mu \nu} = 0
\end{equation}
which reduces to
\begin{eqnarray}
\label{fluc-eq1}
& &\hat{H}_{RS} \hat{\psi}(z) = \frac{m^2}{2} \hat{\psi}(z)
                                                         \\   \nonumber
& &\hat{H}_{RS} = -\frac{1}{2} \partial_z^2 + 
\frac{15}{8(|z| + R)^2} - \frac{3}{2} k \delta(z)
\end{eqnarray}
where $R \equiv 1/k$ is a radius of $AdS_5$ and 
\begin{eqnarray}
\label{defin-1}
& & z = \epsilon(y)
\frac{e^{k|y|} - 1}{k}    \\  \nonumber
& & h_{\mu \nu}(x, y) = e^{-\frac{k}{2} |y|} \hat{\psi}(y) e^{i p x}
                                                  \\  \nonumber
& & m^2 = -p^2.
\end{eqnarray}

At this stage we should stress the fact that if our calculation involves
the brane-bending effect, Eq.(\ref{dasi1}) is modified into
\begin{equation}
\label{dasi2}
\left[ e^{2 k |y|} \Box^{(4)} + \partial_y^2 - 4 k^2 + 4 k \delta(y) \right]
h_{\mu \nu} = - \Sigma_{\mu \nu} \delta(y)
\end{equation}
where $\Sigma_{\mu \nu}$ is a tensor quantity and represents the 
brane-bending effect\cite{garr00}. Thus, the additional term plays the
role of the source and recovers the tensor structure of $h_{\mu \nu}$.
Although it might be possible, in principle, to compute the propagator
of Eq.(\ref{dasi2}) from the viewpoint of SQM with imposing a general 
BC on the brane after computing $\Sigma_{\mu \nu}$ explicitly and 
changing Eq.(\ref{dasi2}) as a Schr\"{o}dinger-type equation, it may lead 
an extreme complication. Since our interest is to examine the effect of the
general boundary condition in the Newton potential,
we will confine ourselves to 
Eq.(\ref{fluc-eq1}) without recovering the tensor structure.

The Newton potential in the bulk is in general computed by the time-integration
of the retarded Green function 
$U(\vec{x}_2, y_2; \vec{x}_1, y_1; t)$\cite{garr00,dvali00}:
\begin{equation}
\label{formu1}
V(\vec{x}_2, y_2; \vec{x}_1, y_1) \equiv
\int dt U(\vec{x}_2, y_2; \vec{x}_1, y_1; t).
\end{equation}
In terms of the time-dependent propagator, 
$U(\vec{x}_2, y_2; \vec{x}_1, y_1; t)$ is represented as
\begin{equation}
\label{formu2}
U(\vec{x}_2, y_2; \vec{x}_1, y_1; t) = \frac{1}{M^3}
\int \frac{d^4p}{(2\pi)^3} e^{-\frac{k}{2} (|y_1| + |y_2|)}
G[y_1 + R, y_2 + R; t] e^{i \vec{p} \cdot (\vec{x}_2 - \vec{x}_1)}
e^{-p_0 t} \delta(p_0 - |\vec{p}|).
\end{equation}
The factor $e^{-\frac{k}{2} (|y_1| + |y_2|)}$ in Eq.(\ref{formu2}) is 
introduced due to the redefinition of $h_{\mu \nu}(x, y)$ in 
Eq.(\ref{defin-1}). The time-integration in Eq.(\ref{formu1}) changes the 
Euclidean propagator $G[y_1 + R, y_2 + R; t]$ into the fixed-energy 
amplitude $\hat{G}[y_1 + R, y_2 + R; \frac{p_0^2}{2}]$ as following:
\begin{equation}
\label{formu3}
V(\vec{x}_2, y_2; \vec{x}_1, y_1) = \frac{1}{8 \pi^3 M^3}
e^{-\frac{k}{2} (|y_1| + |y_2|)} 
\int d^4 p e^{i \vec{p} \cdot (\vec{x}_2 - \vec{x}_1)}
\delta(p_0 - |\vec{p}|) \hat{G}[y_1 + R, y_2 + R; \frac{p_0^2}{2}].
\end{equation}
Thus the $p_0$-integration in Eq.(\ref{formu3}) results in 
\begin{equation}
\label{formu4}
V(\vec{x}_2, y_2; \vec{x}_1, y_1) = \frac{1}{8 \pi^3 M^3}
e^{-\frac{k}{2} (|y_1| + |y_2|)}
\int d^3 \vec{p} e^{i \vec{p} \cdot (\vec{x}_2 - \vec{x}_1)}
\hat{G}[y_1 + R, y_2 + R; \frac{\vec{p}^2}{2}].
\end{equation}
If the brane has three spatial dimensions and is located at $y=0$, the 
Newton potential on the brane is expressed as 
\begin{equation}
\label{general}
V(r) \equiv V(|\vec{x}_2 - \vec{x}_1|, y_1=y_2=0) = 
\frac{1}{2\pi^2 M^3 r}
\int_0^{\infty} dm m \sin mr \hat{G}[R, R; \frac{m^2}{2}]
\end{equation}
where $r \equiv |\vec{x}_2 - \vec{x}_1|$. 

Eq.(\ref{general}) enables us to compute the Newton potential generated by the 
localized gravity on the brane from the fixed-energy amplitude
$\hat{G}[R, R; \frac{m^2}{2}]$. Since the general fixed-energy amplitude which
depends on the BC is computed Ref.\cite{park02-1,park02-2} when there is 
no $4d$ induced Einstein term, we can compute the Newton potential by making 
use of Eq.(\ref{general}). This will be examined in the next section 
in detail.

\section{Newton Potential from Fixed-Energy Amplitude : without $4d$ Induced
Einstein term}
In this section we will compute the Newton potential arising due to the 
localized gravity on the brane when there is no $4d$ induced Einstein term.
Thus what we have to do first is to derive the fixed-energy amplitude for the
linearized gravitational fluctuation (\ref{fluc-eq1}). In 
Ref.\cite{park02-1,park02-2} Eq.(\ref{fluc-eq1}) is slightly generalized
as following
\begin{eqnarray}
\label{slight}
& &\hat{H}_1 \hat{\psi}(z) = E \hat{\psi}(z)    \\   \nonumber
& &\hat{H}_1 = \hat{H}_0 - v \delta(z)      \\   \nonumber
& &\hat{H}_0 = -\frac{1}{2} \partial_z^2 + 
\frac{g}{(|z| + c)^2}.
\end{eqnarray}
Of course, $\hat{H}_1$ in Eq.(\ref{slight}) coincides with $\hat{H}_{RS}$ in 
Eq.(\ref{fluc-eq1}) when $c = 1/k \equiv R$, $g = 15/8$, and 
$v = 3k / 2$.

As we commented before we require the bulk spacetime is a single copy of 
$AdS_5$. This requirement leads us naturally to combine the usual
Schulman procedure\cite{gav86,schul86} for the treatment of the 
$\delta$-function potential with an half-line constraint. The half-line
constraint usually makes the fixed-energy amplitude to be dependent on the
BC on the brane. This procedure is explicitly addressed in 
Ref.\cite{park02-1,park02-2}. The final expression of the fixed-energy
amplitude for $\hat{H}_1$ under these circumstance is 
\begin{eqnarray}
\label{fixed1}
\hat{G}_1[a, b; E]&=&\hat{G}_0^D[a, b; E]      \\   \nonumber
&+& \frac{\sqrt{a b}}{c v}
\frac{K_{\gamma}(\sqrt{2E} a) K_{\gamma}(\sqrt{2E} b)}
     {K_{\gamma}^2(\sqrt{2E} c)}
\left[ \left( \frac{\gamma - \frac{1}{2}}{2 \xi c v} - 1 \right)
      + \frac{\sqrt{2E}}{2 \xi v}
        \frac{K_{\gamma - 1}(\sqrt{2E} c)}
             {K_{\gamma}(\sqrt{2E} c)}
\right]^{-1}
\end{eqnarray}
where $\gamma \equiv \sqrt{1 + 8g} / 2$ and $K_{\gamma}(z)$ is an usual
modified Bessel function. The parameter $\xi$ is explicitly introduecd 
in Eq.(\ref{fixed1}). The real parameter $\xi$ parametrizes the BCs for the
fixed-energy amplitude corresponding to $\hat{H}_0$. For example, 
$\xi = 1$ (or $\xi = 0$) means that we choosed the Neumann (or Dirichlet)
BC for the fixed-energy amplitude corresponding to $\hat{H}_0$.
$\hat{G}_0^D[a, b; E]$ in Eq.(\ref{fixed1}) is a fixed-energy amplitude
for $\hat{H}_0$ when we adopt a Dirichlet BC at the brane. The explicit 
form of $\hat{G}_0^D[a, b; E]$ is given in Ref.\cite{park02-1,park02-2}.
However we do not need to know its explicit form. The only one we should know is
the fact that it satisfies the usual Dirichlet BC, {\it i.e.}
$\hat{G}_0^D[a, c; E] = \hat{G}_0^D[c, b; E] = 0$.

The fixed-energy amplitude on the brane $\hat{G}_1[c, c; E]$ is easily 
computed from Eq.(\ref{fixed1}):
\begin{equation}
\label{g1cce}
\hat{G}_1[c, c; E] = \frac{1}{v}
\left[ \left( \frac{\gamma - \frac{1}{2}}{2 \xi c v} - 1 \right)
      + \frac{\sqrt{2E}}{2 \xi v}
        \frac{K_{\gamma - 1}(\sqrt{2E} c)}
             {K_{\gamma}(\sqrt{2E} c)}
\right]^{-1}.
\end{equation}
Thus inserting Eq.(\ref{g1cce}) into Eq.(\ref{general}) we can explicitly 
compute the Newton potential.

Firstly, we consider $\xi = 1/2$ case which corresponds to the case of usual
RS brane. Letting $c = R$, $g = 15/8$ and $v = 3 / 2 R$ makes the fixed-energy
amplitude to be
\begin{equation}
\label{rsfixed1}
\hat{G}_1^{RS}[R, R; E] = \Delta_0 + \Delta_{KK}
\end{equation}
where
\begin{eqnarray}
\label{d0dkk}
\Delta_{0}&=&\frac{1}{E R}    \\   \nonumber
\Delta_{KK}&=&\frac{1}{\sqrt{2E}}
\frac{K_0(\sqrt{2E} R)}{K_1(\sqrt{2E} R)}.
\end{eqnarray}
Of course, $\Delta_{0}$ and $\Delta_{KK}$ represent the zero mode and KK
excitation respectively. Inserting Eq.(\ref{rsfixed1}) into Eq.(\ref{general})
it is easy to show that the Newton potential on the RS brane reduces to
\begin{equation}
\label{rsnewton1}
V_{RS}(r) = V_{0,RS}(r) + \Delta V_{RS}(r)
\end{equation}
where
\begin{eqnarray}
\label{rsnewton2}
& &V_{0,RS}(r) = \frac{G}{r}    \\   \nonumber
& &\Delta V_{RS}(r) = \frac{G}{\pi} \frac{1}{r}
\int_0^{\infty} du \sin \left( \frac{r}{R} u \right)
\frac{K_0(u)}{K_1(u)}
\end{eqnarray}
and $4d$ Newton constant is defined as $G \equiv 1 / 2 \pi M^3 R$.

It is not difficult to show that the integral in $\Delta V^{RS}(r)$ is not 
well-defined. Thus we need an appropriate regularization. This is achieved 
by introducing the infinitesimal parameter $\epsilon$ as 
following\cite{park03-2,kiri02}
\begin{equation}
\label{rsnewton3}
\Delta V_{RS}(r) = \frac{G}{\pi} \frac{1}{r} J
\end{equation}
where
\begin{equation}
\label{def-J}
J \equiv \lim_{\epsilon \rightarrow 0}
\int_0^{\infty} du \sin \left( \frac{r}{R} u \right)
\frac{K_0(u)}{K_1(u)} e^{-\epsilon u}.
\end{equation}

The factor $\sin \left( \frac{r}{R} u \right)$ in Eq.(\ref{rsnewton3}) is 
crucial to extract the long-range behavior and short-range behavior of the 
Newton potential. Firstly, let us consider the long-range behavior,
{\it i.e.} $r >> R$. The high oscillation behavior of 
$\sin \left( \frac{r}{R} u \right)$ in this case makes the small
$u$ region to be a dominant contribution. Since $K_0(u) \sim -\ln u$ 
and $K_1(u) \sim 1/u$ at $u \sim 0$, $J$ becomes approximately in
this region
\begin{equation}
\label{def-J2}
J \sim - \lim_{\epsilon \rightarrow 0} \int_{0}^{\infty} du
\sin \left( \frac{r}{R}  u \right) u \ln u e^{-\epsilon u}.
\end{equation}
Using an integration formula
\begin{eqnarray}
\label{int-formu1}
& &\int_0^{\infty} dx x e^{-q x} \sin (px) \ln x 
= \frac{1}{p^2 + q^2} \sin \left( 2 \tan^{-1} \frac{p}{q} \right)
\Bigg[ 1 - \gamma - \frac{1}{2} \ln (p^2 + q^2)             \\ \nonumber
& &
\hspace{6.0cm}
        + 
       \tan^{-1} \frac{p}{q} \cot \left(2 \tan^{-1} \frac{p}{q} \right)
\Bigg]
\end{eqnarray}
where $\gamma$ in Eq.(\ref{int-formu1}) is an Euler's constant, it is simple
to show $J \sim \pi R^2 / 2 r^2$.

Secondly, let us consider the short-range behavior, {\it i.e.} $r << R$.
In this case contrary to the long-range behavior the large $u$ region makes
a dominant contribution to $J$. Since at this region 
$K_0(u) \sim K_1(u) \sim \sqrt{\frac{\pi}{2 u}} e^{-u}$, $J$ reduces to 
\begin{equation}
\label{def-J3}
J \sim \int_0^{\infty} du \sin \left(\frac{r}{R} u \right)
e^{-\epsilon u}
\end{equation}
which makes $J$ to be $R/r$. Thus we can summarize the behavior of $J$ as 
following:
\begin{equation}
\label{J-summary}
J \equiv \lim_{\epsilon \rightarrow 0}
\int_0^{\infty} du \sin \left(\frac{r}{R} u \right)
\frac{K_0(u)}{K_1(u)} e^{-\epsilon u}
\sim
\left\{ \begin{array}{ll}
        \frac{\pi R^2}{2 r^2}   & \hspace{1.0cm}  \mbox{if $r >> R$}   \\
        \frac{R}{r}             & \hspace{1.0cm}  \mbox{if $r << R$}.
        \end{array}
\right.
\end{equation}

Eq.(\ref{J-summary}) is confirmed numerically.
Fig. 1 shows that $J$ and $\pi R^2 / 2r^2$ at long-range. For simple
numerical calculation we choosed $\epsilon = 0.003$. Fig. 1 indicates 
that $J$ and $\pi R^2 / 2r^2$ merge with each other at $r \sim 15 R$.
Of course, if we choosed smaller $\epsilon$, the mergence may occur
rapidly. Fig. 2 shows that $J$ and $R / r$ at short-range with
choosing $\epsilon = 0.003$. Fig. 2 also indicates that $J$ and $R / r$
merge with each other at $r \sim 0.05 R$. 

Using Eq.(\ref{J-summary}) one can show that the Newton potential on the 
RS brane becomes
\begin{equation}
\label{rs-newton1}
V_{RS}(r) \sim
\left\{ \begin{array}{ll}
        \frac{G}{r} \left(1 + \frac{R^2}{2 r^2}\right) & \hspace{1.0cm}
                                         \mbox{if $r >> R$}   \\
        \frac{G R}{\pi r^2} \left( 1 + \frac{\pi r}{R} \right) & \hspace{1.0cm}
                                         \mbox{if $r << R$}.
        \end{array}
\right.
\end{equation}
Thus we have $5d$ Newton potential $1/r^2$ at the short-range. Of course, the
Newton potential is $4d$-type $1/r$ at long-range as shown in 
Ref.\cite{rs99-2}. 

Here, we should point out that the long-range behavior of the subleading term
of $V_{RS}(r)$ is different from both Eq.(\ref{rs-newton}) and 
Eq.(\ref{imp-newton}) due to $1/2$ factor. In fact, the subleading term 
for our approach is $J / \pi$ and we computed $J$ with adopting the simple
regularization scheme in Eq.(\ref{def-J}). However, the different scheme 
may assign different value to $J$. In this way, the coefficient of the 
subleading term is dependent on the regularization scheme. It seems to be 
interesting to find a suitable regularization scheme which yields
Eq.(\ref{rs-newton}) or Eq.(\ref{imp-newton}). Since, however, our interest
in this paper is only to find a global behavior of Newton potential 
in general singular
brane, we will not explore this issue further.

Now, we consider $\xi = 1$ case which means that the Neumann BC is chosen.
Inserting $c = R$, $g = 15/8$ and $v = 3 / 2R$ into Eq.(\ref{g1cce}) one
can show that the fixed-energy amplitude on the Neumann brane is 
\begin{equation}
\label{neufixed1}
\hat{G}_1^N[R, R; \frac{m^2}{2}] = 
\frac{\frac{2}{m} \frac{K_2(mR)}{K_1(mR)}}
     {1 - \frac{3}{2} \frac{1}{m R} \frac{K_2(mR)}{K_1(mR)}}.
\end{equation}
Inserting Eq.(\ref{neufixed1}) into (\ref{general}) shows that the Newton
potential arising due to the localized gravity on the Neumann brane 
reduces to 
\begin{equation}
\label{neu-newton1}
V_N(r) = \frac{1}{\pi^2 M^3 R r} \int_0^{\infty} du
\sin \left(\frac{r}{R} u\right)
\frac{\frac{K_2(u)}{K_1(u)}}
     {1 - \frac{3}{2 u} \frac{K_2(u)}{K_1(u)}}.
\end{equation}
Since the term $\sin \left(\frac{r}{R} u\right)$ makes that the small $u$ 
region contributes dominantly at the long-range, the Newton potential at this
range becomes
\begin{equation}
\label{neu-newton2}
V_N(r) \sim \frac{-2}{\pi^2 M^3 R r} \int_0^{\infty} du
\sin \left(\frac{r}{R} u\right) 
\frac{u}{3 - u^2}.
\end{equation}
Since the small $u$ should contribute to $V_N(r)$ dominantly, one can 
approximately change Eq.(\ref{neu-newton2}) into
\begin{equation}
\label{neu-newton3}
V_N(r) \sim \frac{-2}{\pi^2 M^3 R r} \int_0^{\infty} du
\sin \left(\frac{r}{R} u\right) 
\frac{u}{3 + u^2}
\end{equation}
which results in
\begin{equation}
\label{neu-newton4}
V_N(r) \sim - \frac{1}{\pi M^3 R r}
e^{-\sqrt{3} \frac{r}{R}}
\end{equation}
at the long-range, {\it i.e.} $r >> R$. Thus one can conclude that the 
gravitational force at the Neumann brane is exponentially suppressed. It is 
in fact conjectured from the fact that $\hat{G}_1^N[R, R; E]$ has a pole
at $m_N \sim 2.48 R^{-1}$, which makes the graviton on the Neumann brane
to be massive.

At the short-range one can use the asymptotic formula for the modified 
Bessel function, which results in
\begin{equation}
\label{neu-newton5}
V_N(r) \sim \frac{1}{\pi^2 M^3 R r}
\left[\lim_{\epsilon \rightarrow 0} 
\int_0^{\infty} du \sin \left(\frac{r}{R} u\right) e^{-\epsilon u}
- \frac{3}{2}
\int_0^{\infty} du
\frac{\sin \left(\frac{r}{R} u\right)}{\frac{3}{2} - u}  \right]
\end{equation}
where the infinitesimal parameter $\epsilon$ is introduced again for the
regularization.
Carrying out the integration using the integral formula
\begin{equation}
\label{inte-formu2}
\int_0^{\infty} \frac{\sin(a x)}{\beta - x} dx = 
\sin (\beta a) ci (\beta a) - \cos (\beta a) [si (\beta a) + \pi]
\end{equation}
where $si(z)$ and $ci(z)$ are usual sine and cosine integral respectively,
one can express $V_N(r)$ at short-range as
\begin{equation}
\label{neu-newton6}
V_N(r) \sim \frac{1}{\pi^2 M^3 R r}
\left[ \frac{R}{r} - \frac{3}{2}
\left\{\sin \left(\frac{3r}{2R}\right) ci \left(\frac{3r}{2R}\right)
- \cos \left(\frac{3r}{2R}\right) si \left(\frac{3r}{2R}\right) 
- \pi \cos \left(\frac{3r}{2R}\right) \right\} \right].
\end{equation}
Since $r << R$, one can expand $V_N(r)$ using the following expansions of 
$si(z)$ and $ci(z)$\cite{abra72}:
\begin{eqnarray}
\label{expan1}
si(z)&=& - \frac{\pi}{2} + \sum_{k=1}^{\infty}
\frac{(-1)^{k+1} z^{2k-1}}{(2k-1) (2k-1)!}   \\   \nonumber
ci(z)&=&\gamma + \ln z + \sum_{k=1}^{\infty} (-1)^k 
\frac{z^{2k}}{2k (2k)!}
\end{eqnarray}
where $\gamma = 0.577 \cdots$ is an Euler's constant. Then finally 
$V_N(r)$ reduces to 
\begin{equation}
\label{neu-newton7}
V_N(r) \sim \frac{G_5}{r^2}
\left[ 1 - \left(\frac{3r}{2R}\right)^2 \ln \left(\frac{3r}{2R}\right) \right]
\end{equation}
at short-range, where $5d$ Newton constant $G_5$ is defined 
as $G_5 = 1 / \pi^2 M^3$. Thus,
for the Neumann brane $5d$ Newton potential arises at the short-range like
RS brane. The different point, however, is that $V_N(r)$ has a logarithmic 
sub-leading correction.

Now, finally let us consider $\xi = 0$ case which means we have chosen
the Dirichlet BC at the brane for the fixed-energy amplitude of 
$\hat{H}_0$ in Eq.(\ref{slight}). If one naively inserts $\xi = 0$ in 
Eq.(\ref{fixed1}), $\hat{G}_1^D[a, b; E]$ becomes to be identical to 
$\hat{G}_0^D[a, b; E]$ because the modification term in Eq.(\ref{fixed1})
arising due to the $\delta$-function potential via Scuulman
procedure\cite{gav86,schul86} vanishes at $\xi = 0$. Even in this case,
however, one can generate the non-trivial fixed-energy amplitude through a 
coupling constant renormalization. In order to adopt this procedure we should
regard the coupling constant $v$ as an unphysical and infinite bare one. 
Then one can introduce a physical renormalized coupling constant
$v^{ren}$ which is related to $v$ by 
$v^{ren} = (1/v - 2 \epsilon) / 2 \epsilon^2$, where $\epsilon$ is an 
infinitesimal parameter. Through this renormalization procedure one
can derive a non-trivial fixed-energy 
amplitude\cite{park02-1,park02-2}\footnote{there is a factor $2$ mistake
in Eq.(26) of Ref.\cite{park02-1}}:
\begin{equation}
\label{diri-fixed1}
\hat{G}_1^{D, ren}[a, b; \frac{m^2}{2}] = 
\hat{G}_0^D[a, b; E] + 
\frac{2\sqrt{a b}}{\left( R v^{ren} + \frac{3}{2} \right)
                   + m R \frac{K_1(m R)}{K_2(m R)}}
\frac{K_2(m a) K_2(m b)}{K_2^2(m R)}
\end{equation}
which results in
\begin{equation}
\label{diri-fixed2}
\hat{G}_1^{D, ren}[R, R; \frac{m^2}{2}] =
\frac{2 R}{\left( R v^{ren} + \frac{3}{2} \right)
                   + m R \frac{K_1(m R)}{K_2(m R)}}.
\end{equation}
Inserting Eq.(\ref{diri-fixed2}) into Eq.(\ref{general}) the Newton potential
arising due to the exchange of the gravity localized on the Dirichlet
brane becomes
\begin{equation}
\label{diri-newton1}
V_D(r) = \frac{1}{\pi^2 M^3 R r}
\int_0^{\infty} du u \sin \left(\frac{r}{R} u \right)
\frac{1}{\left( R v^{ren} + \frac{3}{2} \right) + u \frac{K_1(u)}{K_2(u)}}.
\end{equation}

Firstly, let us consider a case of $R v^{ren} + 3/2 = 0$. Comparing 
Eq.(\ref{diri-fixed2}) in this case with Eq.(\ref{rsfixed1}) the
fixed-energy amplitude $\hat{G}_1^{D,ren}[R, R; E]$ becomes 
$2 \hat{G}_1^{RS}[R, R; E]$, which implies 
$V_D(r) = 2 V_{RS}(r)$ which is summarized in Eq.(\ref{rs-newton1}).

Next, let us consider a case of $R v^{ren} + 3/2 \neq 0$. Since the 
small $u$ region contributes dominantly in the long-range, 
Eq.(\ref{diri-newton1}) becomes
\begin{equation}
\label{diri-newton2}
V_D(r) \sim \frac{2}{\pi^2 M^3 R r}
\int_0^{\infty} du 
\frac{u \sin \left( \frac{r}{R} u \right)}
     {u^2 + (2 R v^{ren} + 3)},
\end{equation}
which reduces to 
\begin{equation}
\label{diri-newton3}
V_D(r) \sim \frac{1}{\pi M^3 R r}
e^{-\sqrt{2 R v^{ren} + 3} \frac{r}{R}}.
\end{equation}
The expontial suppression of $V_D(r)$ in this long-range behavior indicates
that the gravity localized on the Dirichlet brane is massive.

If $r << R$, the same calculation makes $V_D(r)$ to be
\begin{equation}
\label{diri-newton4}
V_D(r) \sim \frac{1}{\pi^2 M^3 R r}
\left[\lim_{\epsilon \rightarrow 0} \int_0^{\infty} du
      \sin \left(\frac{r}{R} u\right) e^{-\epsilon u}
      - \left(R v^{ren} + \frac{3}{2} \right)
      \int_0^{\infty} du
      \frac{\sin \left(\frac{r}{R} u\right)}
           {u + \left(R v^{ren} + \frac{3}{2} \right)}
\right]
\end{equation} 
where the infinitesimal parameter $\epsilon$ is introduced again for the
regularization. Using the integral formula
\begin{equation}
\label{inte-formu3}
\int_0^{\infty} \frac{\sin a x}{x + \beta} dx = f(a \beta)
\end{equation}
where $f(z) \equiv ci(z) \sin(z) - si(z) \cos(z)$, $V_D(r)$ becomes
\begin{equation}
V_D(r) \sim \frac{1}{\pi^2 M^3 R r}
\left[ \frac{R}{r} - \left(R v^{ren} + \frac{3}{2} \right)
       f\left( \frac{r}{R} \left[R v^{ren} + \frac{3}{2} \right] \right)
                                              \right].
\end{equation}

Thus the short-range behavior of $V_D(r)$ is governed by the remormalized
coupling constant. If $\frac{r}{R} \left(R v^{ren} + \frac{3}{2} \right) >> 1$,
$V_D(r)$ becomes
\begin{equation}
\label{diri-newton5}
V_D(r) \sim \frac{2 R^2 \left(R v^{ren} + \frac{3}{2} \right)^{-2}}
                 {\pi^2 M^3 r^4}
\left[ 1 - \left(R v^{ren} + \frac{3}{2} \right)^{-2} \frac{12 R^2}{r^2}
\right].
\end{equation}
When deriving Eq.(\ref{diri-newton5}) we used the asymptotic 
expression\cite{abra72}:
\begin{equation}
\label{f-expansion}
f(z) = \frac{1}{z} \left( 1 - \frac{2!}{z^2} + \frac{4!}{z^2} - \cdots
                   \right).
\end{equation}
It is interesting to note that the leading order in this case is 
$1/r^4$ which should be a leading term of the seven-dimensional Newton
law. It is unclear at least for us why this pecular behavior of Newton 
potential arises on Dirichlet brane.

If 
$\frac{r}{R} \left(R v^{ren} + \frac{3}{2} \right) << 1$, we should use
Eq.(\ref{expan1}) which results in
\begin{equation}
\label{diri-newton6}
V_D(r) \sim \frac{1}{\pi^2 M^3 r^2}
\left[1 - \frac{\pi}{2} \left(R v^{ren} + \frac{3}{2} \right) \frac{r}{R} 
\right].
\end{equation}
Thus in this case $V_D(r)$ exhibits the $5d$-type Newton potential.

\section{Newton Potential from Fixed-Energy Amplitude : with $4d$ Induced
Einstein term}
In this section we will compute the Newton potential on the brane when the
$4d$ induced Einstein term is involved. The remarkable feature in this case
is an appearance of the $\delta$-function potential which has an 
energy(or mass)-dependent coupling constant in the linearized fluctuation
equation\cite{park03-2,kiri02}:
\begin{eqnarray}
\label{hamil1}
& &\hat{H}_2 \hat{\psi}(z) = E \hat{\psi}(z)   \\   \nonumber
& &\hat{H}_2 = \hat{H}_1 + \lambda E \delta(z)
\end{eqnarray}
where $\lambda \equiv M_4^2 / M^3$ and $\hat{H}_1$ is defined in 
Eq.(\ref{slight}). Thus, the fixed-energy amplitude for 
$\hat{H}_2$ can be obtained by performing the Schulman 
procedure\cite{gav86,schul86} again
\begin{equation}
\label{schul2}
\hat{G}_2[a, b; E] = \hat{G}_1[a, b; E] - 
\frac{\hat{G}_1[a, c; E] \hat{G}_1[c, b; E]}
     {\frac{1}{\lambda E} + \hat{G}_1[c, c; E]}.
\end{equation}
Inserting $a=b=c$ in Eq.(\ref{schul2}) yields the fixed-energy amplitude
on the brane
\begin{equation}
\label{fixed4-1}
\hat{G}_2[c, c; E] = 
\frac{\frac{1}{\lambda E} \hat{G}_1[c, c; E]}
     {\frac{1}{\lambda E} + \hat{G}_1[c, c; E]}.
\end{equation}
Thus, combining Eq.(\ref{general}) and (\ref{fixed4-1}) we can compute the 
Newton potential generated by the exchange of the gravity localized on the
brane. 

Firstly, let us consider $\xi = 1/2$ case in which the fixed-energy 
amplitude reduces by Eq.(\ref{fixed4-1}) to 
\begin{equation}
\label{fixed4-2}
\hat{G}_2^{RS}[R, R; \frac{m^2}{2}] = \frac{2}{m}
\frac{K_2(mR)}{2 K_1(mR) + \lambda m K_2(mR)}.
\end{equation}
Of course, we used $c = R$, $g = 15/8$ and $v = 3 / 2 R$ when deriving 
Eq.(\ref{fixed4-2}). Inserting Eq.(\ref{fixed4-2}) into (\ref{general}) the 
Newton potential becomes 
\begin{equation}
V_{2,RS}(r) = \frac{1}{\pi^2 M^3 R r}
\int_0^{\infty} du \sin \left(\frac{r}{R} u\right)
\frac{K_2(u)}{2 K_1(u) + \frac{\lambda}{R} u K_2(u)}.
\end{equation}

Firstly, let us consider the long-range behavior of $V_{2,RS}(r)$. Since 
small $u$ region contributes dominantly at long-range, one can approximate
\begin{eqnarray}
\label{modi-bsl1}
K_1(u)&\sim&\frac{1}{u} + \frac{u}{2} \ln u     \\   \nonumber
K_2(u)&\sim&\frac{1}{u^2} - \frac{1}{2}
\end{eqnarray}
which makes $V_{2,RS}(r)$ to be 
\begin{eqnarray}
\label{newton4-2}
& &V_{2,RS}(r) \sim
\frac{1}{\pi^2 M^3 R \left(2 + \frac{\lambda}{R} \right) r}
\Bigg[\int_0^{\infty} du \frac{\sin \left(\frac{r}{R} u\right)}{u}
                                                           \\   \nonumber
& &
\hspace{5.0cm}
- \frac{1}{2 + \frac{\lambda}{R}}
 \lim_{\epsilon \rightarrow 0}
\int_0^{\infty} du u e^{-\epsilon u} \sin \left(\frac{r}{R} u\right) \ln u
                                            \Bigg].
\end{eqnarray}
In Eq.(\ref{newton4-2}) we introduced again the infinitesimal parameter
$\epsilon$ for the proper regularizarion. Making use of Eq.(\ref{int-formu1}),
one can show that $V_{2,RS}(r)$ at long-range obeys a $4d$ Newton 
law with $1/r^3$ subleading correction as following:
\begin{equation}
\label{newton4-3}
V_{2,RS}(r) \sim
\frac{1}{2\pi M^3 R \left(2 + \frac{\lambda}{R} \right) r}
\left[1 + \frac{R^2}{\left(2 + \frac{\lambda}{R} \right) r^2} \right].
\end{equation}

Next, let us examine the short-range behavior of $V_{2,RS}(r)$. Since the 
large $u$ region makes a dominant contribution to $V_{2,RS}(r)$, the 
asymptotic formulae
\begin{eqnarray}
\label{modi-bsl2}
K_1(u)&\sim&\sqrt{\frac{\pi}{2 u}} e^{-u} 
\left(1 + \frac{3}{8 u}\right)               \\   \nonumber
K_2(u)&\sim&\sqrt{\frac{\pi}{2 u}} e^{-u}
\left(1 + \frac{15}{8 u}\right)
\end{eqnarray}
can be used. Then one can change $V_{2,RS}(r)$ into
\begin{eqnarray}
\label{newton4-4}
V_{2,RS}(r)&=& \frac{1}{\pi^2 M^3 \lambda r}
\Bigg[\int_0^{\infty} 
      \frac{\sin \left(\frac{r}{R} u\right)}
           {u + \frac{R}{\lambda} \left(2 + \frac{15 \lambda}{8 R} \right)}
       du
                                       \\    \nonumber
&+& \frac{15}{8}
    \frac{1}{\frac{R}{\lambda} \left(2 + \frac{15 \lambda}{8 R} \right)}
\left\{ \int_0^{\infty} \frac{\sin \left(\frac{r}{R} u\right)}{u} du 
       - \int_0^{\infty} 
      \frac{\sin \left(\frac{r}{R} u\right)}
           {u + \frac{R}{\lambda} \left(2 + \frac{15 \lambda}{8 R} \right)}
       du
\right\}           \Bigg].
\end{eqnarray}
Making use of the integral formula (\ref{inte-formu3}), $V_{2,RS}(r)$ at 
$r << R$ reduces to 
\begin{equation}
\label{newton4-5}
V_{2,RS}(r) \sim
\frac{1}{\pi^2 M^3 \lambda r}
\left[ f\left(\frac{2r}{\lambda} + \frac{15r}{8R} \right)
       + \frac{15}{8}
         \frac{1}{\frac{R}{\lambda} \left(2 + \frac{15 \lambda}{8 R}\right)}
       \left\{ \frac{\pi}{2} - 
               f\left(\frac{2r}{\lambda} + \frac{15r}{8R} \right)
       \right\}
\right].
\end{equation}

If the quantum effect of the one loop is so large, {\it i.e.}
$\lambda >> R >> r$, the Newton potential $V_{2,RS}(r)$ becomes
\begin{equation}
\label{newton4-6}
V_{2,RS}(r) \sim
\frac{1}{\pi^2 M^3 \lambda r}
\left[ f\left(\frac{15r}{8R} \right) 
       + \frac{15}{8}
         \frac{1}{\frac{15}{8} + \frac{2 R}{\lambda}}
       \left\{ \frac{\pi}{2} -
              f\left(\frac{15r}{8R} \right) \right\}   \right],
\end{equation}
which reduces to approximately
\begin{equation}
\label{newton4-7}
V_{2,RS}(r) \sim
\frac{1}{2\pi M^3 \lambda r}
\left[1 + \frac{4 r}{\pi \lambda} \ln 
      \left(\frac{15 r}{8 R}\right)  \right].
\end{equation}
Thus, the Newton potential recovers the $4d$-type $1/r$ gravitational 
potential. If the quantum effect of the one loop is very small compared
to $R$, {\it i.e.} $\lambda << R$, Eq.(\ref{newton4-5}) becomes
\begin{equation}
\label{newton4-8}
V_{2,RS}(r) \sim
\frac{1}{\pi^2 M^3 \lambda r}
f \left( \frac{2r}{\lambda} \right).
\end{equation}
Thus in the region $\lambda << r << R$, which means the quantum effect is 
extremely small, the Newton potential exhibits a $5d$-type potential with 
$1/r^4$ correction term as following
\begin{equation}
\label{newton4-9}
V_{2,RS}(r) \sim
\frac{1}{2 \pi^2 M^3 r^2}
\left(1 - \frac{\lambda^2}{2 r^2}\right).
\end{equation}
When deriving Eq.(\ref{newton4-9}) we have used the expansion 
(\ref{f-expansion}). If $r << \lambda << R$, we should use the expansion
(\ref{expan1}), which results in
\begin{equation}
\label{newton4-10}
V_{2,RS}(r) \sim
\frac{1}{2\pi M^3 \lambda r}
\left(1 + \frac{4 r}{\pi \lambda} \ln \frac{2 r}{\lambda} \right).
\end{equation}
Thus at the extremely small-range the $4d$ Newton law is recovered. This
result is also confirmed in Ref.\cite{park03-2,kiri02}. 

In appendix A we examined the Newton potential when $\xi = 1$. Let us 
summarize the result. The long-range behavior in this case is an exponential
suppression like the picture without the $4d$ induced Einstein term:
\begin{equation}
\label{newton4-11}
V_{2,N}(r) \sim - 
\frac{1}{\pi M^3 (R + 2\lambda) r}
exp \left\{-\frac{r}{R} 
          \sqrt{\frac{3}{1 + \frac{2 \lambda}{R}}} \right\}.
\end{equation}
This indicates that the graviton localized on the Neumann brane is massive.
In the region of $\lambda << r << R$ the Newton potential shows the $5d$-type
with $1/r^4$ subleading correction:
\begin{equation}
\label{newton4-12}
V_{2,N}(r) \sim
\frac{1}{\pi^2 M^3 r^2}
\left(1 - \frac{2 \lambda^2}{r^2}\right).
\end{equation}
However, at the extremely short-range, {\it i.e.} $r << \lambda << R$, the 
potential recovers the $4d$-type Newton law:
\begin{equation}
\label{newton4-13}
V_{2,N}(r) \sim
\frac{1}{2 \pi M^3 \lambda r}
\left(1 + \frac{2 r}{\pi \lambda} \ln \frac{r}{\lambda} \right).
\end{equation}

In appendix B we compute the Newton potential at $\xi = 0$. In this case the
fixed-energy amplitude as well as the Newton potential depend on the 
renormalized coupling constant $v^{ren}$ as in the case without the $4d$
induced Einstein term. If $R v^{ren} + \frac{3}{2} = 0$, the final expression
of the Newton potential is related to $V_{2,RS}(r)$ as following:
\begin{equation}
\label{newton4-14}
V_{2,D}(r) = 2 V_{2,RS}(r) \bigg|_{\lambda \rightarrow 2 \lambda}.
\end{equation}
 
If $R v^{ren} + \frac{3}{2} \neq 0$, Newton potential is exponentially 
suppressed in the long-range as the case in the absence of the $4d$
induced Einstein term as following:
\begin{equation}
\label{newton4-15}
V_{2,D}(r) \sim 
\frac{1}{2\pi M^3 \left(\lambda + \frac{R}{2} \right) r}
exp \left\{- \frac{r}{R} \sqrt{\frac{R v^{ren} + \frac{3}{2}}{\frac{1}{2}
                                + \frac{\lambda}{R}}} \right\}.
\end{equation}
This indicates that the graviton localized on the Dirichlet brane is massive.
In the short-range the Newton potential becomes
\begin{equation}
\label{newton4-16}
V_{2,D}(r) \sim
\frac{1}{\pi^2 M^3 r^2}
\left(1 - \frac{2 \lambda^2}{r^2}\right)
\end{equation}
when $\lambda << r << R$ and 
\begin{equation}
\label{newton4-17}
V_{2,D}(r) \sim
\frac{1}{2\pi M^3 \lambda r}
\left(1 + \frac{2 r}{\pi \lambda} \ln \frac{r}{\lambda} \right)
\end{equation}
when $r << \lambda << R$. Thus, we have an intermediate range
$\lambda << r << R$ in which the $5d$ Newton law emerges. 
It is interesting to note that $V_{2,D}(r)$ at short-range is independent of
the renormalized coupling constant $v^{ren}$. Furthermore, the potential on
Dirichlet brane coincides with that on Neumann brane in the short-range.

\section{Conclusion}
In this paper we examined Newton law generated by the localized gravity on the
general singular brane when the bulk spacetime is a single copy of 
$AdS_5$. We used a fixed-energy amplitudes which are obtained from the
linearized gravitational fluctuation equation by applying the technique
of SQM. Since the bulk is a single copy of $AdS_5$, SQM naturally makes the
fixed-energy amplitude to be dependent on the BC at the location of the 
brane. We examined Newton potential on RS brane($\xi = 1/2$), Neumann
brane($\xi = 1$), and Dirichlet brane($\xi = 0$). 

For RS brane the usual RS scenario gives rise to the $4d$-type Newton potential
at long-range and $5d$-type at short-range. However, the $4d$ Einstein term
induced by a quantum effect of one-loop changes the general behavior of 
Newton potential. In this case there is an intermediate range
$\lambda << r << R$, in which Newton potential is five-dimensional.
At other ranges the four-dimensional Newton law is recovered.

For Neumann brane the long-range Newton potential is exponentially suppressed
regardless of the existence of the $4d$ induced Einstein term. This is because
the gravity localized on the Neumann brane acquires a mass. In the short-range
Newton potential becomes five-dimensional with a logarithmic subleading 
correction if there is no $4d$ induced Einstein term. The Einstein term in
Neumann brane yields a similar intermediate range to that in RS brane, where
Newton law is five-diemnsional.

For Dirichlet brane the non-trivial fixed-energy amplitude can be 
derived {\it via} the coupling constant renormalization scheme. Thus, 
Newton potential on the Dirichlet brane is usually dependent on the 
renormalized coupling constant $v^{ren}$. If $v^{ren} = -3 / 2R$, the 
final expression of Newton potential is proportional to that in 
RS brane. If $v^{ren} \neq -3 / 2R$, the long-range behavior of Newton
potential is expontially suppressed due to the massive gravity regardless of
the $4d$ induced Einstein term. In short-range Newton potential exhibits an
pecular $1/r^4$ behavior if 
$\frac{r}{R} \left(R v^{ren} + \frac{3}{2} \right) >> 1$. However, the 
$4d$ Einstein term makes the short-range behavior of Newton potential
on Dirichlet brane to be exactly identical to that on Neumann brane.

Recently, Newton law arising due to the localized gravity on various 
curved brane is examined when bulk is $dS_5$ or $AdS_5$\cite{noji02}. 
Our SQM technique may be applied straightforwardly to these various
scenarios. The explicit result will be given elsewhere.

\vspace{1cm}

{\bf Acknowledgement}:  
This work was supported by the Kyungnam University
Research Fund, 2002.

\newpage
\begin{appendix}{\centerline{\bf Appendix A}}
\setcounter{equation}{0}
\renewcommand{\theequation}{A.\arabic{equation}}
In this appendix we will examine the Newton potential arising due to the 
gravity localized on the Neumann brane when the $4d$ induced Einstein term
exists. The fixed-energy amplitude for $\xi=1$ with $c = R$, $g = 15/8$ and 
$v = 3 / 2R$ yields 
\begin{equation}
\label{A-fixed1}
\hat{G}_2^N[R, R; \frac{m^2}{2}] = 
\frac{\frac{2}{m}  \frac{K_2(m R)}{K_1(m R)}}
     {1 + \left( \lambda m - \frac{3}{2 m R} \right) \frac{K_2(m R)}{K_1(m R)}}
\end{equation}
and the corresponding Newton potential derived from Eq.(\ref{general}) is 
\begin{equation}
\label{A-newton1}
V_{2,N}(r) = \frac{1}{\pi^2 M^3 R r}
\int_0^{\infty} du \sin \left(\frac{r}{R} u \right)
\frac{\frac{K_2(u)}{K_1(u)}}
     {1 + \left(\frac{\lambda}{R} u - \frac{3}{2 u} \right) 
       \frac{K_2(u)}{K_1(u)}}.
\end{equation}
If two unit masses are separated at long distance compared to the 
$AdS_5$ radius, {\it i.e.} $r >> R$, the small $u$ region contributes to 
$V_{2,N}(r)$ dominantly. Thus we can use an expansion (\ref{modi-bsl1}) for the
modified Bessel function, which results in approximately
\begin{equation}
\label{A-newton2}
V_{2,N}(r) \sim - \frac{2}{\pi^2 M^3 (R + 2 \lambda) r}
            \int_0^{\infty} du \sin \left(\frac{r}{R} u \right)
\frac{u}{\frac{3}{1 + \frac{2 \lambda}{R}} + u^2}.
\end{equation}
Making use of the integral formula
\begin{equation}
\label{A-int1}
\int_0^{\infty} 
\frac{x \sin(a x)}{\beta^2 + x^2} dx = \frac{\pi}{2} e^{- a \beta},
\end{equation}
one can show that the gravitational force on the Neumann brane is 
exponentially suppressed as following:
\begin{equation}
\label{A-newton3}
V_{2,N}(r) \sim - \frac{1}{\pi M^3 (R + 2 \lambda) r}
exp \left\{ - \frac{r}{R}
            \sqrt{\frac{3}{1 + \frac{2 \lambda}{R}}} \right\}.
\end{equation}

If two unit masses are separated at short distance, the large $u$ region 
contributes dominantly. Thus we can use the asymptotic formula 
(\ref{modi-bsl2}), which results in the following Newton potential
\begin{eqnarray}
\label{A-newton4}
& &V_{2,N}(r) \sim \frac{1}{\pi^2 M^3 \lambda r}
\Bigg[ \int_0^{\infty} du
\frac{\sin \left(\frac{r}{R} u \right)}
     {u + \left(\frac{15}{8} + \frac{R}{\lambda} \right)}
                                                     \\   \nonumber
& & 
\hspace{4.0cm}
+ \frac{\frac{15}{8}}
     {\frac{15}{8} + \frac{R}{\lambda}}
\left\{\int_0^{\infty} du 
\frac{\sin \left(\frac{r}{R} u \right)}{u}
- \int_0^{\infty} du
\frac{\sin \left(\frac{r}{R} u \right)}
     {u + \left(\frac{15}{8} + \frac{R}{\lambda} \right)}
                                                     \right\}
                                                     \Bigg].
\end{eqnarray}
Computing the integrations of Eq.(\ref{A-newton4}) with use of 
Eq.(\ref{inte-formu3}), one can express $V_{2,N}(r)$ at short-range as
\begin{equation}
\label{A-newton5}
V_{2,N}(r) \sim \frac{1}{\pi^2 M^3 \lambda r}
\left[ f\left( \frac{15r}{8R} + \frac{r}{\lambda} \right)
     + \frac{\frac{15}{8}}{\frac{15}{8} + \frac{R}{\lambda}}
     \left\{ \frac{\pi}{2} - f\left( \frac{15r}{8R} + \frac{r}{\lambda} \right)
     \right\}            \right].
\end{equation}
If $\lambda << R$, $V_{2,N}(r)$ in Eq.(\ref{A-newton5}) becomes
\begin{equation}
\label{A-newton6}
V_{2,N}(r) \sim \frac{1}{\pi^2 M^3 \lambda r}
f\left( \frac{15r}{8R} + \frac{r}{\lambda} \right).
\end{equation}
Thus using the expansion of the sine and cosine integral functions
(\ref{expan1}) and (\ref{f-expansion}), the Newton potential on the 
Neumann brane becomes
\begin{equation}
\label{A-newton7}
V_{2,N}(r) \sim \frac{1}{\pi^2 M^3 r^2}
\left(1 - \frac{2 \lambda^2}{r^2}\right)
\end{equation}
in the region of $\lambda << r << R$, and
\begin{equation}
\label{A-newton8}
V_{2,N}(r) \sim \frac{1}{2 \pi M^3 \lambda r}
\left(1 + \frac{2 r}{\pi \lambda} \ln \frac{r}{\lambda} \right)
\end{equation}
in the region of $r << \lambda << R$. 
\end{appendix}

\vspace{1cm}
\begin{appendix}{\centerline{\bf Appendix B}}
\setcounter{equation}{0}
\renewcommand{\theequation}{B.\arabic{equation}}
In this appendix we will examine the Newton potential arising due to the 
gravity localized on the Dirichlet brane when the $4d$ induced Einstein term
exists. The fixed-energy amplitude for $\xi = 0$ case can be obtained from
Eq.(\ref{fixed4-1}) with replacing $\hat{G}_1[c, c; E]$ by a renormalized
fixed-energy amplitude $\hat{G}_1^{ren}[c, c; E]$. Then using 
Eq.(\ref{diri-fixed2}) one can easily calculate the fixed-energy
amplitude
\begin{equation}
\label{B-fixed1}
\hat{G}_2^D[R, R; \frac{m^2}{2}] = 
\frac{2R}
     {\left(R v^{ren} + \frac{3}{2}\right) + mR \frac{K_1(mR)}{K_2(mR)}
      + R \lambda m^2}
\end{equation}
which leads the Newton potential to 
\begin{equation}
\label{B-newton1}
V_{2,D}(r) = \frac{1}{\pi^2 M^3 R r}
\int_0^{\infty} du
\frac{u \sin \left(\frac{r}{R} u\right)}
     {\left(R v^{ren} + \frac{3}{2}\right) + u \frac{K_1(u)}{K_2(u)} + 
      \frac{\lambda}{R} u^2}.
\end{equation}

Firstly, let us examine the case of $R v^{ren} + \frac{3}{2} = 0$. Comparing 
Eq.(\ref{B-fixed1}) with Eq.(\ref{fixed4-2}) one can conclude
\begin{equation}
\label{B-fixed2}
\hat{G}_2^D[R, R; \frac{m^2}{2}] =
2 \hat{G}_2^{RS}[R, R; \frac{m^2}{2}] \bigg|_{\lambda \rightarrow 2 \lambda}.
\end{equation}
Thus, the Newton potential $V_{2,D}(r)$ in this case can be read from 
$V_{2,RS}(r)$ as following:
\begin{equation}
\label{B-newton2}
V_{2,D}(r) = 2 V_{2,RS}(r) \bigg|_{\lambda \rightarrow 2 \lambda}.
\end{equation}

Next, let us examine the case of $R v^{ren} + \frac{3}{2} \neq 0$. Using the 
expansion (\ref{modi-bsl1}) one can show $V_{2,D}(r)$ reduces to in the
long-range, {\it i.e.} $r >> R$,
\begin{equation}
\label{B-newton3}
V_{2,D}(r) \sim
\frac{1}{\pi^2 M^3 \left(\lambda + \frac{R}{2}\right) r}
\int_0^{\infty} du
\frac{u \sin \left(\frac{r}{R} u\right)}
     {\frac{R v^{ren} + \frac{3}{2}}{\frac{1}{2} + \frac{\lambda}{R}} + u^2},
\end{equation}
which exhibits an exponential suppression
\begin{equation}
\label{B-newton4}
V_{2,D}(r) \sim 
\frac{1}{2\pi M^3 \left(\lambda + \frac{R}{2} \right) r}
exp \left\{- \frac{r}{R} \sqrt{\frac{R v^{ren} + \frac{3}{2}}{\frac{1}{2}
                                + \frac{\lambda}{R}}} \right\}.
\end{equation}
In the short-range we should use the expansion (\ref{modi-bsl2}) which makes
$V_{2,D}(r)$ to be 
\begin{eqnarray}
\label{B-newton5}
& &V_{2,D}(r) \sim
\frac{1}{\pi^2 M^3 \lambda r}
\Bigg[ \int_0^{\infty} du
\frac{\sin \left(\frac{r}{R} u \right)}
     {u + \left(\frac{15}{8} + \frac{R}{\lambda} \right)}
                                                     \\   \nonumber
& & 
\hspace{4.0cm}
 + \frac{\frac{15}{8}}
     {\frac{15}{8} + \frac{R}{\lambda}}
\left\{\int_0^{\infty} du 
\frac{\sin \left(\frac{r}{R} u \right)}{u}
- \int_0^{\infty} du
\frac{\sin \left(\frac{r}{R} u \right)}
     {u + \left(\frac{15}{8} + \frac{R}{\lambda} \right)}
                                                     \right\}
                                                     \Bigg].
\end{eqnarray}
It is interesting to note that $V_{2,D}(r)$ is independent of the renormalized
coupling constant $v^{ren}$. Furthermore, $V_{2,D}(r)$ in Eq.(\ref{B-newton5})
is exactly same with $V_{2,N}(r)$ in Eq.(\ref{A-newton4}). Thus the short-range
behavior of $V_{2,D}(r)$ should be same with that of $V_{2,N}(r)$.

\end{appendix}

\begin{figure}
\caption{Plot of $J$ and $\pi R^2 / 2 r^2$ with choosing $\epsilon = 0.003$. 
This figure indicates 
$J$ and $\pi R^2 / 2 r^2$ merge with each other at $r \sim 15 R$. Thus the 
long-range behavior of $J$ in Eq.(\ref{J-summary}) is confirmed.}  
\end{figure}
\vspace{0.4cm}
\begin{figure}
\caption{Plot of $J$ and $R / r$ with choosing $\epsilon = 0.003$.
This figure indicates
$J$ and $R / r$ merge with each other at $r \sim 0.05 R$. Thus the
short-range behavior of $J$ in Eq.(\ref{J-summary}) is confirmed.} 
\end{figure}

\newpage
\epsfysize=20cm \epsfbox{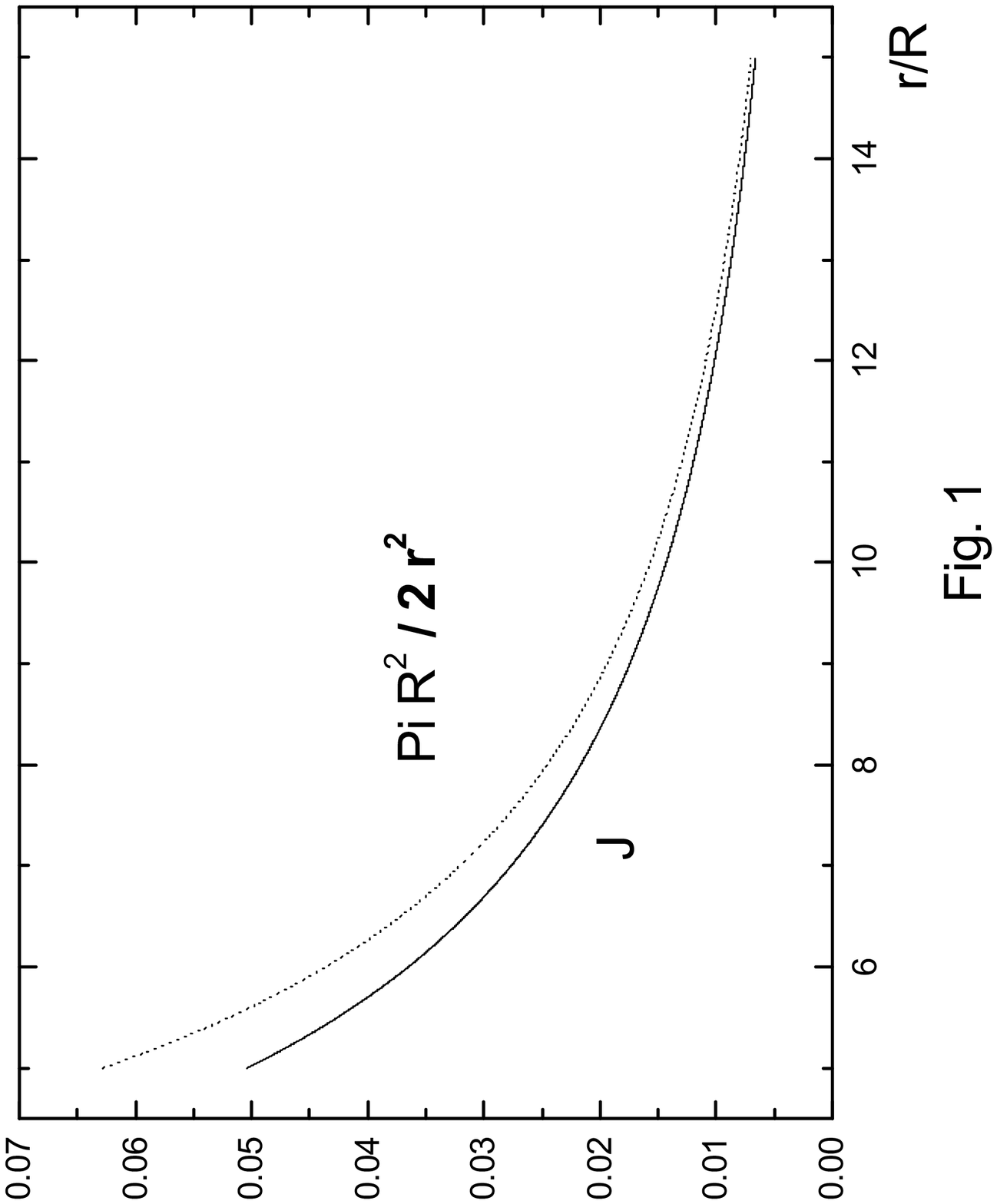}
\newpage
\epsfysize=20cm \epsfbox{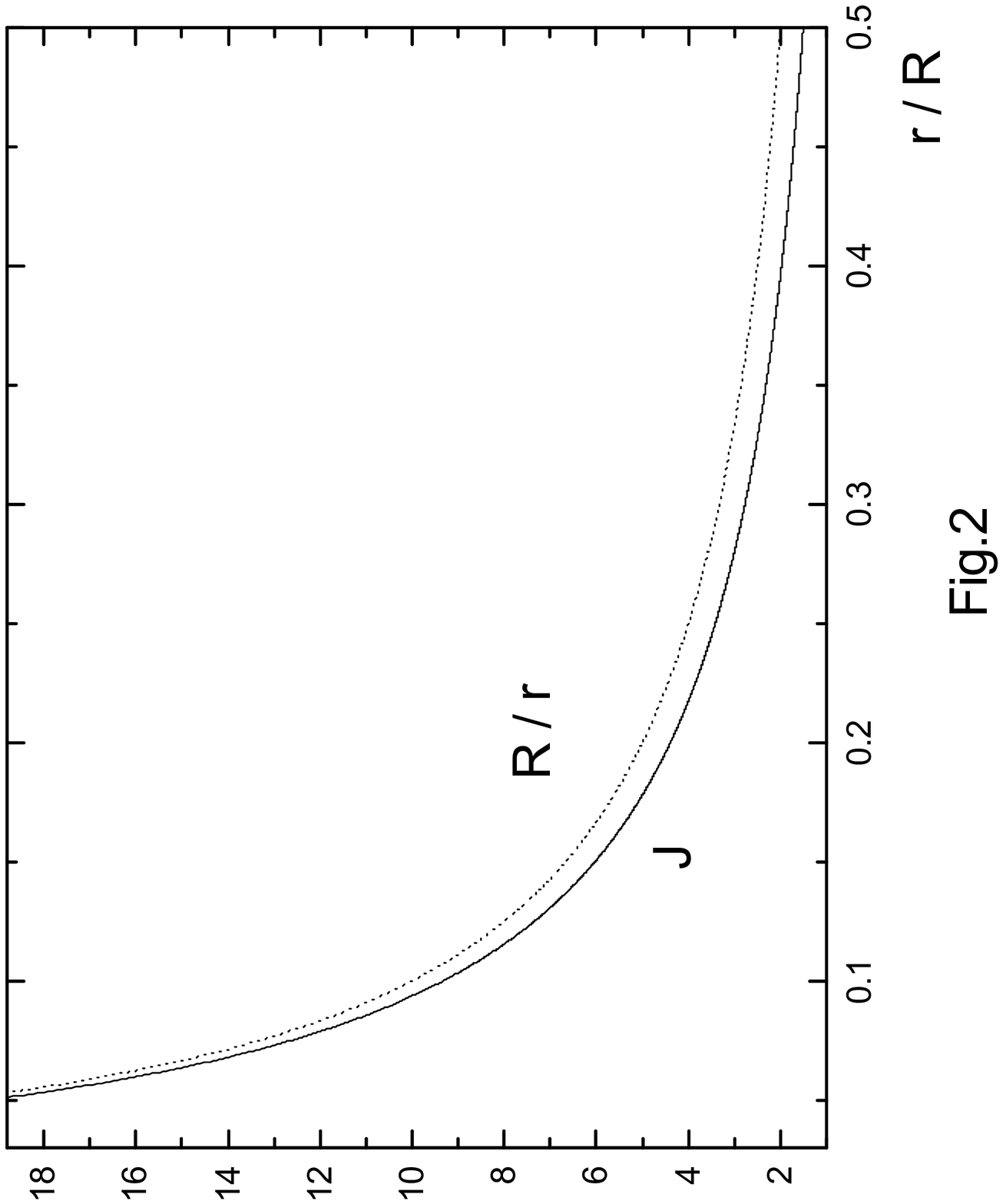}
\end{document}